\documentclass[sigconf]{acmart}

\AtBeginDocument{%
  \providecommand\BibTeX{{%
    \normalfont B\kern-0.5em{\scshape i\kern-0.25em b}\kern-0.8em\TeX}}}

\setcopyright{acmcopyright}
\copyrightyear{2018}
\acmYear{2018}
\acmDOI{XXXXXXX.XXXXXXX}

\acmConference[]{WWW}{June 03--05,
  2018}{Woodstock, NY}
\acmPrice{15.00}
\acmISBN{978-1-4503-XXXX-X/18/06}

\usepackage{booktabs}
\usepackage{algorithm}
\usepackage{algorithmic}
\usepackage{enumitem}
\usepackage{xcolor}
\usepackage{subfigure}
\usepackage{multicol}
\usepackage{multirow}
\usepackage{algorithm}
\usepackage{algorithmic}
\usepackage{makecell}
\usepackage{graphicx}
\usepackage{caption}
\usepackage{subfigure}
\usepackage{xspace}
\newcommand{\modelname}{FINED\xspace}

\begin{document}

\title{FINED: Feed Instance-Wise Information Need with Essential and Disentangled Parametric Knowledge from the Past}


\author{Kounianhua Du}
\affiliation{%
  \institution{Shanghai Jiao Tong University}
  \city{Shanghai}
  \country{China}}
\email{774581965@sjtu.edu.cn}

\author{Jizheng Chen}
\affiliation{%
  \institution{Shanghai Jiao Tong University}
  \city{Shanghai}
  \country{China}}
\email{humihuadechengzhi@sjtu.edu.cn}

\author{Jianghao Lin}
\affiliation{
  \institution{Shanghai Jiao Tong University}
  \city{Shanghai}
  \country{China}}
\email{chiangel@sjtu.edu.cn}

\author{Menghui Zhu}
\email{zhumenghui1@huawei.com}
\affiliation{
  \institution{Huawei Noah's Ark Lab}
  \city{Shanghai}
  \country{China}
}

\author{Bo Chen}
\affiliation{%
  \institution{Huawei Noah's Ark Lab}
  \city{Shanghai}
  \country{China}}
\email{chenbo116@huawei.com}


\author{Shuai Li}
\affiliation{%
  \institution{Shanghai Jiao Tong University}
  \city{Shanghai}
  \country{China}}
\email{shuaili8@sjtu.edu.cn}

\author{Yong Yu}
\affiliation{%
  \institution{Shanghai Jiao Tong University}
  \city{Shanghai}
  \country{China}}
\email{yyu@apex.sjtu.edu.cn}

\author{Weinan Zhang}
\affiliation{%
  \institution{Shanghai Jiao Tong University}
  \city{Shanghai}
  \country{China}}
\email{wnzhang@sjtu.edu.cn}

\renewcommand{\shortauthors}{Kounianhua Du et al.}

\begin{abstract}

Recommender models play a vital role in various industrial scenarios, while often faced with the catastrophic forgetting problem caused by the fast shifting data distribution, e.g., the evolving user interests, click signals fluctuation during sales promotions, etc. To alleviate this problem, a common approach is to reuse knowledge from the historical data. However, preserving the vast and fast-accumulating data is hard, which causes dramatic storage overhead. Memorizing old data through a parametric knowledge base is then proposed, which compresses the vast amount of raw data into model parameters. Despite the flexibility, how to improve the memorization and generalization capabilities of the parametric knowledge base and suit the flexible information need of each instance are challenging. In this paper, we propose \modelname to \textbf{F}eed \textbf{IN}stance-wise information need with \textbf{E}ssential and \textbf{D}isentangled parametric knowledge from past data for recommendation enhancement. Concretely, we train a knowledge extractor that extracts knowledge patterns of arbitrary order from past data and a knowledge encoder that memorizes the arbitrary order patterns, which serves as the retrieval key generator and memory network respectively in the following knowledge reusing phase. The whole process is regularized by the proposed two constraints, which improve the capabilities of the parametric knowledge base without increasing the size of it. The essential principle helps to compress the input into representative vectors that capture the task-relevant information and filter out the noisy information.
The disentanglement principle reduces the redundancy of stored information and pushes the knowledge base to focus on capturing the disentangled invariant patterns. 
These two rules together promote rational compression of information for robust and generalized knowledge representations. Extensive experiments on two datasets justify the effectiveness of the proposed method.

\end{abstract}



\received{20 February 2007}
\received[revised]{12 March 2009}
\received[accepted]{5 June 2009}

\maketitle

\section{Introduction}
\label{sec:intro}
Recommender systems play an important role in alleviating information overload and helping users discover relevant content in today's vast digital landscape. They are widely used in various industries, including e-commerce \citep{Amazon}, entertainment \citep{entertainment}, social media \cite{news}, and online streaming platforms \citep{18din, 19dien}.

However, the data distribution shifts fast in recommender systems, e.g., the evolving user interests, the fluctuation of click signals during sale promotions, etc. This fast shifting distribution often results in the catastrophic forgetting problem that models cannot remember old knowledge well. A line of methods is to reuse knowledge from old data in a non-parametric way \citep{20sim}. However, preserving the vast and fast-accumulating raw data is heavy, which causes dramatic storage overhead. Compressing the knowledge from raw data into parameters for reusing is then proposed \citep{qin2024d2k, wang2023data}.

\begin{figure}[t!]
    \centering
    \includegraphics[width=0.48\textwidth]{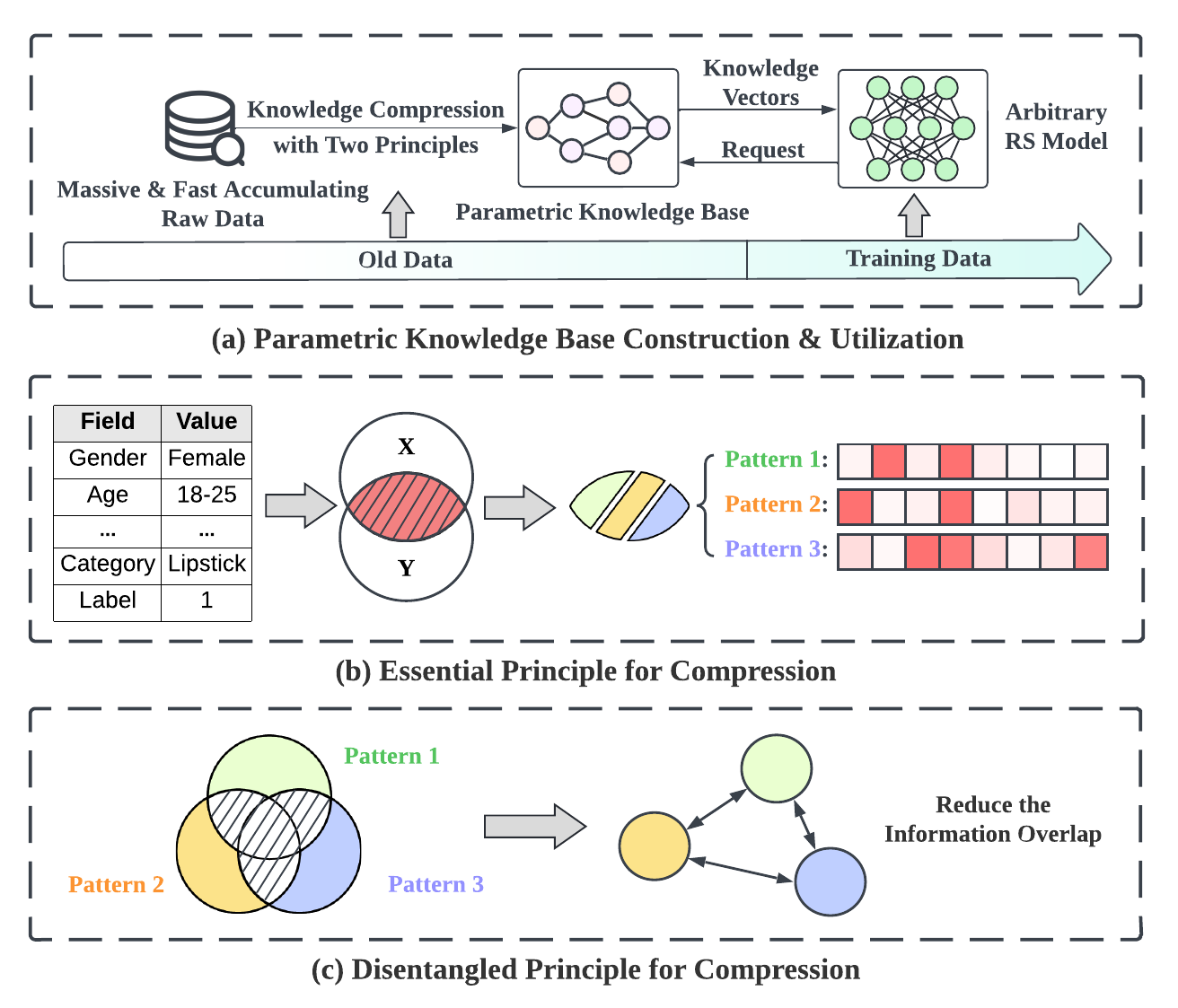}
    \vspace{-10pt}
    \caption{(a) Overall process. (b) The essential principle for compression, which encourages a compressed representation that captures the fundamental knowledge of data and filters out the noises. (c) The disentangled principle for compression, which reduces the redundancy of stored patterns and decomposes the invariance for better generalization.}
    \label{fig:intro}
    \vspace{-10pt}
    
\end{figure}



Despite the flexibility, how to improve the memorization and generalization capabilities of the parameters is challenging due to the noisy information and the diverse patterns existing in the vast amount of old data.
For example, a user named $Alice$ might follow the trend to buy a product $Meal\_Replacement$ during sales promotions and give a positive feedback due to the return cash offer, which is not an actual interest of hers. $Alice-Meal\_Replacement$ forms a spurious pattern that could bias the model from memorizing the invariant pattern. In addition, the patterns existing in the data space are diverse and complex, which are hard to be fully memorized. Some patterns may entangle with each other, leading to the redundant storage and poorer generalization. For example, $Female-Actress-Lipstick$, $Female-Lipstick$, and $Actress-Lipstick$ entangles with each other, where $Actress$ information contains $Female$ information so that $Female-Actress-Lipstick$ leads to redundant memorization. Decomposing $Female-Actress-Lipstick$ into $Female-Lipstick$ and $Actress-Lipstick$ helps to reduce redundancy and meanwhile improve the expressiveness of the features, since the minimal information that has better invariance is kept. 



To solve the challenges above, we propose the essential and disentangled parametric knowledge base, where two constraints regularize a parametric knowledge base for robust and generalized knowledge preserving without increasing the size of the knowledge base.
Concretely, the \textbf{Essential} principle aims to capture the task-relevant information (the red part in Figure~\ref{fig:intro}.(b)) and filter out the noisy information (the grey part in Figure~\ref{fig:intro}.(b)) in the data, which follows the Information Bottleneck principle \citep{ibp} that finds the optimal trade-off between compression and prediction. By striking a balance between these two aspects, the principle enables the model to preserve sufficient information of data while keeping the representation minimal and invariant \citep{achille2018emergence}, which contributes to lower storage cost and higher generalization capability.
The \textbf{Disentanglement} principle eliminates redundant information among different patterns (the grey part in Figure~\ref{fig:intro}.(c)) and decomposes the invariance of different patterns. This reduction of redundancy helps to improve memorization within limited model parameters. And at the same time, as depicted in \citep{montero2020role}, improving disentanglement may contribute to the generalization. 
The two rules together contribute to compressed and rational knowledge vectors that capture the fundamental knowledge of past data and have good generalization ability for future distribution. 

Concretely, our method can be split into two stages: knowledge compression and knowledge utilization. 
During the first stage, we compress the massive old data documents into the learned parameters. Specifically, we first extract informative patterns from the old data using a knowledge extractor. The extracted patterns are then encoded and memorized by a permutation-equivariant encoder for knowledge vectors. The whole process is under the regularization of the essential and disentangled principles. After the compression, the knowledge extractor and knowledge encoder together make up the parametric knowledge base, which serve as the retrieval key generator and memory network in the following stage.
During the knowledge utilization stage, we extract knowledge from the constructed knowledge base for each data instance and utilize it as a supplement for recommendation enhancement. The extensive experiment results validate the effectiveness of the proposed method, as well as its superior model compatibility.

Our contributions can be summarized into three-folds.
\begin{itemize}[leftmargin=10pt]
    \item We propose a parametric knowledge base that compresses the massive old data documents into robust and generalized parametric knowledge, which is a model-agnostic method that could offer diverse knowledge for various base models.
    \item We propose two rules to effectively regularize the parametric knowledge base, which promotes robustness and generalization by extracting essential and disentangled  knowledge from the old data without increasing the size of it.
    \item The knowledge extraction and utilization of \modelname is  instance-level, which offers fine-grained information tailored for each instance for enhancement.
\end{itemize}

Major experiments over different baseline models and ablation studies validate the effectiveness of the proposed method.

\section{Related Work}
\label{sec:re}
\subsection{Conventional Recommendation Backbones}
Conventional recommenders capture collaborative signals based on feature interaction learning and user behavior modeling. 

For feature interaction learning, there are plenty of works mining the interactive patterns of categorical data.
FM \citep{rendle2010factorization} captures feature interactions by inner products. FFM \citep{juan2016field}  further develops field-aware interactions by multiple embeddings settings. NeuFM \citep{he2017neural} proposes to use deep neural networks, improving FM by incorporating a multi-layer perceptron. Wide \& Deep \citep{cheng2016wide} combines the strengths of linear models and DNNs for memorization and generalization respectively. DeepFM \citep{guo2017deepfm} utilizes an FM layer to replace the wide component in \cite{cheng2016wide} to model the pairwise feature interactions. xDeepFM \citep{lian2018xdeepfm} introduces the cross layer to replace the wide component and constructs limited high order feature interactions. PNN \citep{18pnn} proposes a product layer to better model the ``AND'' relations between features. DCN \citep{dcn} learns both low-dimensional feature crossing and high-dimensional nonlinear features efficiently. AutoInt \citep{autoint} learns high-order feature interactions via the self-attention mechanism. 

User behavior modeling plays a crucial role in recommendation systems by providing valuable insights into user preferences, interests, and behaviors. By understanding how users interact with a system and make choices, recommendation algorithms can effectively personalize and tailor recommendations to individual users. DIN \citep{18din} models user interests with attention mechanism. It reduces the influence of historical behavior information of commodities that are not related to the currently estimated advertisement on the current click estimation judgment. DIEN \citep{19dien} further points out that user interests evolve over time and modeled the evolving interests of users with the GRU module. DSIN \citep{dsin} splits different sessions and better models the evolving property of user interests. SIM \citep{20sim} proposes to model long-term user behaviors and aggregates them using the fast SimHash algorithm.

\vspace{-10pt}
\subsection{Distribution Robust Recommenders}
To deal with the fast-shifting distribution of recommendation datasets, various methodologies are came up, including continual learning methods, memory networks, and distributionally robust models. 

Continual learning methods aim to continuously adapt a recommendation model to new data without forgetting previously learned knowledge, which need to adapt to users' changing preferences, new items, or evolving contexts without the computational overhead of retraining from scratch. The main paradigms of continual learning can be summarized into regularization-based methods and methods using modular architecture. 
Regularization-based methods prevent the catastrophic forgetting of old data when the model learns from new data. They work by adding a regularization term to the loss function to constrain the updates on the model parameters \citep{lopez2017gradient, chaudhry2018efficient, mi2020ader,wang2020practical}, thus keeping them close to previous values unless warranted by strong evidence from new data. Methods that use modular architecture grow the base model when faced with newly arrived data \citep{riemer2018learning}. 

Memory networks in recommendation scenarios allow the system to store user interactions and context in a structured memory, enabling the model to refer to past interactions when generating recommendations. 
NMRN \citep{wang2018neural} proposes a streaming recommender with external memories.
RUM \citep{chen2018sequential} introduces a first-in-first-out writing mechanism, which stores and updates users' historical records explicitly for better future prediction.
KSR \citep{huang2018improving} integrates RNN-based networks with Key-Value Memory Networks and utilizes extensive knowledge base information to better encode user preferences. 
HPMN \citep{ren2019lifelong} propose a hierarchical periodic memory network model for lifelong sequential modeling in user response prediction, which can handle long-term sequential patterns and multi-scale sequential patterns of user interests.
MIMN \citep{pi2019practice} presents a co-design solution for handling long sequential user behavior data, combining a separate module called UIC (User Interest Center) in the serving system to manage user interests in real-time and a novel memory-based architecture to capture user interests from long sequential behavior data.

Distributionally Robust Optimization (DRO) in the context of recommendation systems is a powerful method aimed at improving the robustness and stability of recommendations under various types of data uncertainty and modeling assumptions.  SML \citep{zhang2020retrain} employs a neural network-based transfer component, enabling the retraining of recommender systems to focus on new data efficiently and effectively, thereby addressing overfitting and improving recommendation accuracy compared to full model retraining.
InvCF \citep{zhang2023invariant}  generates unbiased preference representations resistant to changes in item popularity, demonstrating the ability to outperform leading models in terms of popularity generalization across multiple datasets and evaluation settings without requiring prior knowledge of popularity distributions.

\section{Preliminaries}
\subsection{Problem Formulation}
\label{sec:pre}
Click-Through-Rate prediction predicts the signal of a user clicking a candidate item, which is an essential and important task in recommender systems. Conventional CTR prediction methods \citep{rendle2010factorization, juan2016field, he2017neural, cheng2016wide, guo2017deepfm, lian2018xdeepfm, autofis, autoint, 18pnn, lin2023map, dcn} can be formulated as 
\begin{equation}
    p(y|X,\theta),
\end{equation}
where $X=\left[x_1, x_2, \dots, x_F\right]$ is the input consisting of features from $F$ feature fields and $\theta$ is the model parameter.
These methods make prediction based solely on target feature and model parameters. Another thread of methods that models user behavior patterns \citep{18din, 19dien, mimn, dsin, 20sim} shows important impacts on making personalized recommendation and brings great performance gains, which can be formulated as 
\begin{equation}
    p(y|X,D_{his},\theta),
\end{equation}
where $D_{his}$ represents the user behavior histories.
These methods focus on modeling user historical pattern by directly inputting users' historical behaviors.

In this paper, we aim to extract diverse patterns existing in old data and build an efficient paramatric knowledge base to supplement the target prediction. The framework can be formulated as 
\begin{align}
    KB_{\theta*} \leftarrow D_{old},\\
    p(y|X,KB_{\theta*},\theta),\\
    p(y|X,D_{his},KB_{\theta*},\theta),
\end{align}
where $D_{old}$ denotes the old data and $KB_{\theta*}$ denotes the knowledge base we construct based on old data.

\subsection{Model Invariance \& Generalization}
Due to the limited capacity of model parameters, improving the quality of the memorized knowledge is important. 

As stated in \citep{achille2018emergence}, an ideal representation should be sufficient, minimal, invariant, and maximally disentangled. It also proves that a sufficient representation of the data is invariant if and only if it contains the smallest amount of information. 
In other words, achieving the optimal balance of sufficiency and minimalist will at the same time promotes invariance, which contributes to the generalization \citep{deng2022strong}. This aligns with the information bottleneck theory \citep{IB} that a model should extract the most relevant information of an input sample $X$ corresponding to $Y$:
\begin{equation}
    \min \left(I(X;S)-\beta I(S;Y)\right),
\end{equation}
where $S$ denotes the representation of the encoded input.

As for the disentanglement requirement, it is proved in machine learning tasks that disentanglement contributes to generalization \citep{montero2020role, yang2023vector}. A common thread of methods to improve the disentanglement is to minimize the mutual information of the representations. 

While mutual information being hard to estimate, one can decrease the upper bound of mutual information to minimize it and increase the lower bound of mutual information to maximize it. CLUB \citep{club} introduces a contrastive log-ratio upper cound of mutual information, which can be represented as
\begin{equation}
\begin{split}
    I(X;Y)&\leq I_{CLUB}(X;Y)\\
    &= E_{p(x,y)}\left[\log p(y|x)\right]-E_{p(x)}E_{p(y)}\left[\log p(y|x)\right].
\end{split}
\end{equation}
When the conditional distribution $p(y|x)$ is not known, one could use a variational distribution $q_\theta(y|x)$ to approximate it and the upper bound then becomes
\begin{equation}
    I_{vCLUB}=E_{p(x,y)}\left[\log q_\theta(y|x)\right]-E_{p(x)}E_{p(y)}\left[\log q_\theta(y|x)\right].
\end{equation}

MINE \citep{mine} proposes a lower bound of the mutual information based on the Donsker-Varadhan representation of KL divergence:
\begin{equation}
\begin{split}
    I(X;Y)&:=D_{KL}(\mathbb{J}||\mathbb{M})\geq \hat{I_w}^{(DV)}(X;Y) \\
    &:=E_\mathbb{J}\left[T_w(x;y)\right] - \log E_\mathbb{M}\left[e^{T_w(x,y)}\right] ,
\end{split}
\end{equation}
where $\mathbb{J}$ is the joint distribution, $\mathbb{M}$ is the product of marginal distributions, and $T_w$ is a discriminator. Then DIM \citep{dim} points out that we do not necessarily need to obtain the precise value of MI and use the Jensen-Shannon Divergence to estimate it instead, where a GAN-style loss is proposed:
\begin{equation}
    L=E_{\mathbb{J}}\left[\log T_w(x,y)\right] + E_{\mathbb{M}}\left[\log(1-T_w(x,y))\right].
\end{equation}
\begin{figure*}[t!]
    \centering
    \includegraphics[width=1.0\textwidth]{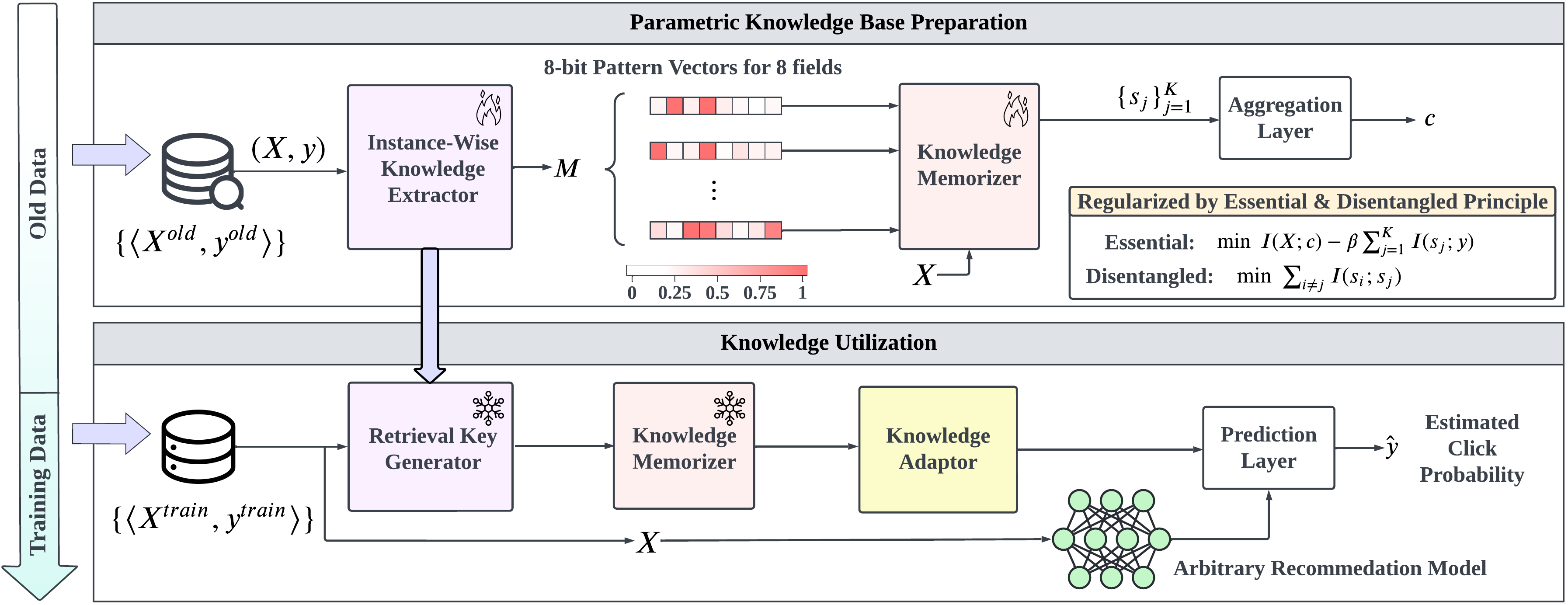}
    \vspace{-15pt}
    \caption{
    The framework of \modelname. In the knowledge compression stage, we compress the essential and disentangled knowledge within old data into the parametric knowledge base. Concretely, we extract patterns within data instances with the knowledge extractor and memorize them through the knowledge encoder that could deal with inputs of arbitrary scales. The overall knowledge compression process is regularized by two principles: essential and disentangled for better generalization and robustness. During prediction, the target could access the frozen knowledge base for instance-wise knowledge and adapt the knowledge to inject it into arbitrary recommendation backbone for enhanced prediction.}
    \label{fig: framework}
    \vspace{-10pt}
\end{figure*}

\vspace{-5pt}
\section{Methodology}
Our framework can be divided into two stages: \textbf{knowledge compression} and \textbf{knowledge utilization}. In the knowledge compression stage, we compress the essential and disentangled knowledge within the old data into compact vectors stored in the parameters of the parametric knowledge base. This stage consists of two modules: a knowledge extractor that extracts informative patterns from the raw data features and a knowledge encoder that encodes and memorizes patterns of arbitrary scales. The compression process is guided by the essential and disentangled principles for robust and generalized knowledge within limited model parameters. After the compression, the knowledge extractor and knowledge encoder made up the parametric knowledge base, which serve as the retrieval key generator and the memory network in the following phase respectively.
In the knowledge utilization stage, we access the knowledge base trained in the previous stage and adapt the knowledge for enhanced prediction.

\subsection{Knowledge Compression}
During this stage, we compress the knowledge of old data into a parametric knowledge base consisting of a knowledge extractor $f(\cdot)$ and a knowledge encoder $g(\cdot)$, guided by the essential and disentangled principles. 
The overall knowledge compression process can be formulated as 
\begin{equation}
    \underbrace{X\xrightarrow[]{f(\cdot)}\{\Tilde{\mathbf{s_j}}\}\xrightarrow[]{g(\cdot)}\{\mathbf{s_j}\}\xrightarrow[]{Aggregator}{\mathbf{c}}}_{\text{Regularized by Essential \& Disentangled}},
\end{equation}
where $X$ denotes a sample of the old data, $\{\Tilde{\mathbf{s_j}}\}$ denotes the different patterns of the input, $\{\mathbf{s_j}\}$ denotes the encoded knowledge vectors, and $\mathbf{c}$ denotes the final abbreviated representation of the input. For simplicity, we omit the footnote of index for data sample. After training, the knowledge extractor $f(\cdot)$ and knowledge encoder $g(\cdot)$
constitute the parametric knowledge base into which the vast old data documents are compressed.

\subsubsection{Essential \& Disentangled Principles}
The whole compression process is regularized by the essential and the disentangled principles.
The essential principle encourages 
 compressed representation that captures the sufficient and minimal information relevant to the task. This elimination of irrelevant or noisy information helps the parametric knowledge base focus only on the invariant patterns and avoids it from being biased by the spurious patterns.
 The disentangled principle decomposes the entanglement of the memorized knowledge patterns, which eliminates the redundancy of knowledge memorizing and promotes the disentanglement of different underlying factors, which helps the parametric knowledge base generalize its knowledge to new instances or distributions more effectively. The formulations of the two principles are given as:

\begin{itemize}[leftmargin=10pt]
    \item \textbf{Essential. $\min I(X;\{\mathbf{s_j}\}) -\alpha I(\{\mathbf{s_j}\};y)$.} The generated knowledge representations should contain as much of the task-relevant information of the original input and as less of task-irrelevant information. 
    It helps to distill data into a sufficient and minimal representation, enabling better generalization and robustness.
    \item \textbf{Disentangled. $\min \sum_{i\ne j} I(\mathbf{s_i};\mathbf{s_j})$}. The generated knowledge representations should contain different aspects of information about the input. It helps to reduce memory redundancy and disentangle underlying factors for better invariance.
\end{itemize}

Together, the two principles contribute to sufficient, minimal, and invariant knowledge representations, which improves the expressiveness and generalization of the parameters of the knowledge base without increasing the size of it. The objective of the two principles can be formulated as 
\begin{equation}
    \min I(X;\{{\mathbf{s_j}}\}) -\alpha I(\{\mathbf{s_j}\};y)+\beta\sum_{i\neq j}I(\mathbf{s_i};\mathbf{s_j}),
\end{equation} 
where $\alpha$ and $\beta$ are hyperparameters to scale the loss components.

Due to the complexity of computing mutual information among a set of variables, we can relax the above objective and obtain the following loss:
\begin{equation}
    l_{reg} =  \underbrace{- \alpha\sum_j I(\mathbf{s_j};y)+ I(X;\mathbf{c})}_{\text{Essential}} + \underbrace{\beta\sum_{i\neq j}I(\mathbf{s_i};\mathbf{s_j})}_{\text{Disentangled}},
    \label{eq:reg}
\end{equation}
where $\mathbf{c}$ is the abbreviated representation of $X$.

To achieve the essential objective, we need to 1) maximize the mutual information between each encoded knowledge vector and the label and at the same time 2) minimize the mutual information between the abbreviated representation and the input. These two strike a balance between compression and prediction. 
For maximizing the mutual information between each encoded knowledge vector and the label, the difficulty lies in that the label space is discrete and only contains two values. Therefore, we follow DIM~\citep{dim} to maximize the distance between the joint distribution and the marginal distribution. Concretely, we regard a pattern representation and a random embedded input with the same label as the joint distribution. For the marginal distribution, we sample the pair as the pattern representation and a random embedded input with the opposite label. 
\begin{equation}
    \begin{split}
    \max \sum_j I(\mathbf{s_j};y) &= \max \sum_j(log D_\theta(\mathbf{{s_j}}, \mathbf{c^+}) \\
    &+ (1 - log D_\theta(\mathbf{{s_j}}, \mathbf{c^-}))).
    \end{split}
\end{equation}
For minimizing the mutual information between the abbreviated representation and the input, we minimize the lower bound of the mutual information proposed in \cite{club}.
\begin{equation}
    \min I(X;\mathbf{c}) = \min I_{vCLUB}(X;\mathbf{c}).
\end{equation}

To achieve the disentangled objective, we minimize the mutual information among encoded knowledge vectors. Since the combinations of different vectors is polynomial, it is hard to train a mutual information estimator for each pair of vectors. Instead, we use a loss that is equivalent to the contrastive loss, which enlarges the discrepancy among different patterns.
\begin{equation}
    \min \sum_{i\neq j}I(\mathbf{s_i};\mathbf{s_j})= \min - \sum_{i}log\frac{z(\mathbf{s_i})z(\mathbf{s_i})}{\sum_{i\neq j}z(\mathbf{s_i})z(\mathbf{s_j})},
\end{equation}
where $z$ is the augmentation operator defined by 
\begin{equation}
    z=MLP(Norm(Dropout(\cdot))). 
\end{equation}

\subsubsection{Knowledge Extractor $f(\cdot)$}
A sample of the data could be represented by $X=\left[x_1, x_2, \dots, x_F\right]$, where $x_i$ is the feature value of the $i$-th field and $F$ is the number of feature fields. For each sample of the old data, we generate different masks to obtain different patterns and encode the patterns for knowledge. Concretely, for an arbitrary input $X$, we use a global attentive readout to extract the mutual relations among features and obtain the context-aware features for mask generation. 

Firstly, we embed the features of input $X=\left[x_1, x_2, \dots, x_F\right]$ by 
\begin{align}
    &\forall x_i \in X, \text{\quad} \mathbf{x}_i=\Phi_1(x_i),\\
    &\mathbf{H_0}=\left[\mathbf{x}_1,\dots,\mathbf{x}_F\right],
\end{align}
where $\Phi_1$ denotes the embedding operation used to generate embeddings for knowledge extraction, $\mathbf{H_0}\in R^{F\times d}$, and $d$ is the embedding size.

Then we feed the embedded input into a global self-attention layer to obtain the context-aware features for patterns extraction.
\begin{equation}
    \mathbf{H} \xleftarrow{GlobalAtt} \mathbf{H_0}.
\end{equation}

After the global self-attention readout, we generate the mask for $K$ patterns of the input by 
\begin{equation}
\label{eq:mask}
    \mathbf{M}=\mathbf{H}\mathbf{P},
\end{equation}
where $\mathbf{M}\in R^{F\times K}$ is the generated mask, $\mathbf{P}\in R^{d\times K}$ is a learnable parameter, and $K$ denotes the number of knowledge patterns we generate.

To make each entry of the mask approximately binary and ensure the gradient consistency, we map each entry $m_{ij}$ of the generated mask matrix $\mathbf{M}$ using the hard concrete distribution \citep{hard}, which is a continuous relaxation of discrete random variables. Concretely, we obtain the masks as
\begin{align}
    &\text{\qquad \quad}u_{ij} \sim U(0,1), \\
    m_{ij}=\sigma&\left(\left(\log u_{ij} - \log(1-u_{ij})+\log m_{ij}\right)/\beta\right),\\
    &m_{ij}=Tanh\left(m_{ij}(\delta-\gamma)+\gamma\right),
\end{align}
where $u_{ij}$ is sampled from a uniform distribution and $\sigma$ denotes the sigmoid function. 

After we obtain the masks, we apply them on the original input to generate different patterns. 
\begin{align}
    \mathbf{H'} &= \left[\Phi_2(x_1),\dots,\Phi_2(x_F)\right],\\
    \Tilde{\mathbf{s}}_j &= \mathbf{H'}\odot \mathbf{M}_{:,j}, \text{\quad} \forall j\in \{1,\dots, K\},
    \label{eq:key}
\end{align}
where $\Phi_2$ is the embedding operation used to generate embeddings for knowledge memorization, $\odot$ denotes the element-wise product, and $\Tilde{\mathbf{s}}_j\in R^{F\times d}$ is a pattern vector.

\subsubsection{Knowledge Encoder $g(\cdot)$}
After we obtain the extracted patterns $\{\Tilde{\mathbf{s_j}}\}$, we feed the patterns into the knowledge encoder to memorize them. Since the generated patterns are of arbitrary lengths, the encoder should be capable of handling inputs of different lengths. Hence, we adopt the classical self-attention architecture~\citep{VaswaniSPUJGKP17}.
 
For every masked knowledge pattern, each entry of it attends to others through a self-attention layer as follows:
\begin{align}
\mathbf{Q} = \mathbf{\Tilde{s}W_Q}, \text{ }\mathbf{K} &= \mathbf{\Tilde{s}W_K}, \text{ }\mathbf{V} = \mathbf{\Tilde{s}W_V}, \\
\mathbf{A} = so&ftmax\left(\frac{\mathbf{QK^T}}{\sqrt{d}}\right),\\
Attention&(\mathbf{Q, K, V}) = \mathbf{AV}.
\end{align}
We represent the above process with $\psi(\cdot)$. Then the encoded vectors for each knowledge pattern and the final abbreviated representation $\mathbf{c}$ of an input could be obtained by:
\begin{align}
    \mathbf{s_j} &= MLP(\psi(\mathbf{\Tilde{s_j}})),\\
    \mathbf{c} &= MLP(AGG_{j}\{\mathbf{{s_j}}\}).
\end{align} 

\subsubsection{Objective}
The training objective of the knowledge compression stage can then be summarized as 
\begin{equation}
    l_{compression} = l_{ce} + \lambda_1 l_{reg} + \lambda_2 l_0,
\end{equation}
where $l_{ce}$ denotes the cross entropy loss of predicting labels on $\mathbf{c}$, $l_{reg}$ corresponds to the two principles proposed in Equation~\eqref{eq:reg}, and $l_0$ denotes the regularization of the generated masks. 

\subsection{Knowledge Utilization}
After we establish the parametric knowledge base $KB_{\theta*} = g(f(\cdot))$ using the old data, we could extract the essential and disentangled knowledge vectors for target prediction enhancement. 
When making prediction for a target sample $X$, we could extract the knowledge from the knowledge base and append it for prediction. The process can be formulated as:
\begin{equation}
    \hat{y} = \Phi_{base}\left(MLP\left(KB_{\theta*} (X)\right), X\right),
\end{equation}
where $\Phi_{base}$ denotes the backbone model and $KB_{\theta*}$ denotes the frozen parametric knowledge base. 

The final prediction is optimized by the cross-entropy loss:
\begin{equation}
l_{pred} = \sum_{\langle X, y \rangle\in D_{train}} (y \log \hat{y} + (1- y) \log(1-\hat{y})).
\end{equation}



\section{Experiments} 
In this section, we empirically evaluate the proposed model with two datasets on the Click-Through-Prediction task. We evaluate the proposed method starting from the following research questions.
\begin{itemize}
    \item[\textbf{RQ1}] Can the proposed model beat the baseline models?
    \item[\textbf{RQ2}]
    Is the proposed parametric knowledge base model-agnostic? Does \modelname offer performance gains for different backbones?
    \item[\textbf{RQ3}] Are the components of the parametric knowledge effective? Do the proposed constriants improve the quality of the parametric knowledge? How does the number of knowledge vectors per data instance influence the performance?
    \item[\textbf{RQ4}] Are the knowledge patterns well captured? Are the objectives of the proposed constraints (i.e., essential \& disentangled principles) achieved during training?
\end{itemize}

\subsection{Experimental Setup}
We use two datasets to validate the proposed model. 
\vspace{-10pt}
\begin{table}[!h]
    \centering
    \caption{Statistics of the Used Datasets.}
    \vspace{-10pt}
    \label{tab:data}
    \resizebox{0.99\linewidth}{!}{
    \begin{tabular}{c|ccccc}
    \toprule
    Dataset & \# Users & \#Items & \#Documents &\# Fields &\# Features\\
    \midrule
      AD & 1,061,768 &827,009 &25,029,435 &12 &3,029,333\\
      Eleme &5,782,482 &1,853,764 &49,114,930 &17 & 16,516,885\\
    \bottomrule
    \end{tabular}}
\end{table}
\vspace{-10pt}

\begin{itemize}[leftmargin=10pt]
\item \textbf{Ali Display\footnote{https://tianchi.aliyun.com/dataset/56}} is a dataset provided by Alibaba to estimate the click-through rate of Taobao display ads.
\item \textbf{Eleme\footnote{https://tianchi.aliyun.com/dataset/131047}} is a dataset provided by Eleme that offered take-away service to users. 
\end{itemize}

The detailed descriptions for the used datasets are summarized in Table~\ref{tab:data}. For the two datasets, the clicked samples are treated as positive samples and the exposed but not clicked samples are treated as negative samples. We split the data using the global timestamps~\citep{20ubr, 21rim}.
For regular baseline models, we use the logs before $T_1$ for training, 50\% of the logs after $T_1$ for validation, and 50\% of the logs after $T_1$ for test. For those supplemented with extra knowledge vectors, we use the logs before $T_0$ for building the knowledge base, the logs between $[T_0, T_1)$ for training base models, 50\% of the logs after $T_1$ for validation, and 50\% of the logs after $T_1$ for test. ($T_0<T_1$). 


We compare the proposed \modelname with four groups of baselines.
\begin{itemize}[leftmargin=10pt]
    \item \textbf{(Group 1) Memory networks}. HPMN \citep{ren2019lifelong} and MIMN \citep{pi2019practice} are memory networks that utilize external memory to preserve additional knowledge.
    \item \textbf{(Group 2) Incremental learning methods}. ADER \citep{mi2020ader} and IncCTR \citep{wang2020practical} continually update the model parameters to adapt to different distributions.
    \item \textbf{(Group 3) Distributionally robust models}. SML \citep{zhang2020retrain} and InvCF \citep{zhang2023invariant} are distributionally robust models that explore invariant patterns for shifting distributions.
    \item \textbf{(Group 4) Parametric knowledge base model}. D2K \citep{qin2024d2k} proposes to use model parameters learned on old data to construct knowledge base for future prediction.
\end{itemize}

\begin{table*}[!t]
\centering
\caption{Comparisons with different baselines. Rel.Impr. denotes the relative AUC improvement of \modelname against each baseline. The symbol * indicates statistically significant improvement with p-value $<0.001$.}
    \label{tab:main_exp}
    \vspace{-10pt}
\begin{tabular}{cc|ccc|ccc}
\hline
\multicolumn{2}{c|}{\multirow{2}{*}{Model}}                                                 & \multicolumn{3}{c|}{AD}       & \multicolumn{3}{c}{Eleme}     \\ \cline{3-8} 
\multicolumn{2}{c|}{}                                                                       & AUC    & Logloss & Rel. Impr. & AUC    & Logloss & Rel. Impr. \\ \hline
\multicolumn{1}{c|}{\multirow{2}{*}{Memory Network}}                 & HPMN                 & 0.6210 & 0.1970  & 2.11\%     & 0.5984 & 0.0963  & 9.68\%     \\
\multicolumn{1}{c|}{}                                                & MIMN                 & 0.6178 & 0.1984  & 2.64\%     & 0.5924 & 0.0987  & 10.79\%    \\ \hline
\multicolumn{1}{c|}{\multirow{2}{*}{Incremental Learning}}           & ADER                 & 0.6241 & 0.1944  & 2.20\%     & 0.6122 & 0.0908  & 7.20\%     \\
\multicolumn{1}{c|}{}                                                & IncCTR               & 0.6194 & 0.1955  & 2.97\%     & 0.6033 & 0.0910  & 8.79\%     \\ \hline
\multicolumn{1}{c|}{\multirow{2}{*}{Distributionally Robust Models}} & SML                  & 0.6221 & 0.1952  & 1.93\%     & 0.5857 & 0.0927  & 12.05\%    \\
\multicolumn{1}{c|}{}                                                & InvCF                & 0.6233 & 0.1945  & 2.33\%     & 0.5961 & 0.0911  & 10.10\%    \\ \hline
\multicolumn{1}{c|}{\multirow{2}{*}{Parametric Knowledge Base}}      & D2K                  & 0.6341 & 0.1940  & 0.58\%     & 0.6401 & 0.1343  & 2.53\%     \\ \cline{2-8} 
\multicolumn{1}{c|}{}                                                & \modelname                  & 0.6378 & 0.1935  & -          & 0.6563 & 0.0881  & -          \\\hline
\end{tabular}
\end{table*}
\begin{table*}[t!]
\centering
\caption{Compatibility Study. 
N/A denotes the original recommendation backbone. Rel.Impr denotes the relative AUC improvement of \modelname against each backbone model. The symbol * indicates statistically significant improvement with p-value $<0.001$.}
    \label{tab:compact}
    \vspace{-10pt}
\begin{tabular}{cc|ccc|ccc}
\toprule
\multicolumn{2}{c|}{\multirow{2}{*}{Model}}                   & \multicolumn{3}{c|}{AD}                                  & \multicolumn{3}{c}{Eleme}                                \\ \cline{3-8} 

\multicolumn{2}{c|}{}                                         & AUC    & Logloss                     & Rel. Impr.        & AUC    & Logloss                     & Rel. Impr.        \\ \hline

\multicolumn{1}{c|}{\multirow{2}{*}{DeepFM}}  & N/A & 0.6220 & \multicolumn{1}{c|}{0.1949} & \multirow{2}{*}{2.54\%} & 0.6221 & \multicolumn{1}{c|}{0.0907} & \multirow{2}{*}{5.36\%} \\ \cline{2-4} \cline{6-7}

\multicolumn{1}{c|}{}                         & w/ \modelname  & 0.6378* & \multicolumn{1}{c|}{0.1935*} &                   & 0.6555*& \multicolumn{1}{c|}{0.0901*} &                   \\ \hline

\multicolumn{1}{c|}{\multirow{2}{*}{DCN}}     & N/A & 0.6231 & \multicolumn{1}{c|}{0.1948} & \multirow{2}{*}{2.25\%} & 0.6386 & \multicolumn{1}{c|}{0.0889} & \multirow{2}{*}{2.77\%} \\ \cline{2-4} \cline{6-7}

\multicolumn{1}{c|}{}                         & w/ \modelname  & 0.6371* & \multicolumn{1}{c|}{0.1936*} &                   & 0.6563* & \multicolumn{1}{c|}{0.0881*} &                   \\ \hline

\multicolumn{1}{c|}{\multirow{2}{*}{PNN}}     & N/A & 0.6299 & \multicolumn{1}{c|}{0.1944} & \multirow{2}{*}{1.03\%} & 0.6324 & \multicolumn{1}{c|}{0.0884} & \multirow{2}{*}{3.61\%} \\ \cline{2-4} \cline{6-7}

\multicolumn{1}{c|}{}                         & w/ \modelname  & 0.6364* & \multicolumn{1}{c|}{0.1935*} &                   &    0.6552*   & \multicolumn{1}{c|}{0.0881*}       &                   \\ \hline

\multicolumn{1}{c|}{\multirow{2}{*}{xDeepFM}} & N/A & 0.6291 & \multicolumn{1}{c|}{0.1943} & \multirow{2}{*}{1.08\%} & 0.6466 & \multicolumn{1}{c|}{0.0883} & \multirow{2}{*}{1.48\%} \\ \cline{2-4} \cline{6-7}

\multicolumn{1}{c|}{}   & w/ \modelname  & 0.6359* & \multicolumn{1}{c|}{0.1939*} &                   &       0.6562* & \multicolumn{1}{c|}{0.0887*}       &                   \\ \hline

\multicolumn{1}{c|}{\multirow{2}{*}{AutoInt}}     & N/A & 0.6339 & \multicolumn{1}{c|}{0.1944} & \multirow{2}{*}{0.35\%} & 0.6464 & \multicolumn{1}{c|}{0.0934} & \multirow{2}{*}{1.39\%} \\ \cline{2-4} \cline{6-7}

\multicolumn{1}{c|}{}                         & w/ \modelname  & 0.6361* & \multicolumn{1}{c|}{0.1937*} &                   &   0.6554*     & \multicolumn{1}{c|}{0.0885*}       &                   \\ \hline

\multicolumn{1}{c|}{\multirow{2}{*}{DIN}}     & N/A & 0.6233 & \multicolumn{1}{c|}{0.1947} & \multirow{2}{*}{2.23\%} & 0.6411 & \multicolumn{1}{c|}{0.0883} & \multirow{2}{*}{2.26\%} \\ \cline{2-4} \cline{6-7}

\multicolumn{1}{c|}{}                         & w/ \modelname  & 0.6372* & \multicolumn{1}{c|}{0.1935*} &                   &   0.6556*     & \multicolumn{1}{c|}{0.0883*}       &                   \\ \hline
\end{tabular}
\end{table*}

Since our method is model-agnostic, we also test its compatibility with different backbone models. 
We evaluate the compatibility on six widely used conventional recommendation backbones, the feature interaction operators of which include dnn-based (xDeepFM \citep{lian2018xdeepfm}, DCN \citep{dcn}), product-based (DeepFM \citep{guo2017deepfm}, PNN \citep{18pnn}), and attention-based (DIN \citep{18din}, AutoInt \citep{autoint}) methodologies. 

The evaluation metrics include area under ROC curve (AUC) and negative log-likehood (LogLoss). For the hyperameters, details can be found in Section~\ref{sec:hyper} in the Appendix.

\subsection{Main Results (RQ1-RQ2)}
The overall performance of the proposed method is displayed in Table~\ref{tab:main_exp}. And the compatibility experiment on different backbones is displayed in Table~\ref{tab:compact}.

(\textbf{RQ1}) For the model performance, from Table~\ref{tab:main_exp}, one can see that \modelname consistently outperform all the baseline models, including memory networks, incremental learning methods, distributionally robust models, and the recent proposed parametric knowledge base model. The improvements are statistically significant under p-value $<0.001$. This validates that compressing knowledge of old data documents into parameters could preserve useful knowledge and offer performance gains for recommender systems. 

(\textbf{RQ2}) From Table~\ref{tab:compact}, one can see that \modelname consistently offers significant performance improvements for all the selected backbone models. The improvements are statistically significant under p-value $<0.001$. This shows that \modelname has superior model compatibility, which demonstrates the effectiveness of our established parametric knowledge base under the essential and disentangled constraints.

\subsection{Ablation Study (RQ3)}
\subsubsection{Impact of the Two Constraints}
To enforce the parametric knowledge base to memorize the essential and disentangle knowledge patterns within limited model parameters, we propose two constraints to regularize the knowledge compression process. In this section, we investigate the effectiveness of the two proposed constraints. We conduct experiments with backbones on which \modelname achieves the best performance, which is DeepFM for AD and DCN for eleme, respectively. We remove the regularizations of the essential and disentangled principles to see the impacts of them. 
\begin{table}[h]
    \centering
    \caption{Impacts of the two constraints.}
    \vspace{-10pt}
    \label{tab:cons}
    \begin{tabular}{c|cccc}
    \hline
    Model & \multicolumn{2}{c}{AD} & \multicolumn{2}{c}{Eleme}\\\hline
    \modelname                  & 0.6378 & 0.1935           & 0.6563 & 0.0881\\
        \modelname w/o Disentangled & 0.6356 & 0.1939          & 0.6433 & 0.0886          \\
                                            \modelname w/o Essential    & 0.6347 & 0.1937          & 0.6520 & 0.0884       \\
                                              \modelname w/o Both         & 0.6304 & 0.1940        & 0.6411 & 0.0888        \\ \hline
    \end{tabular}
\end{table}                                           
From Table~\ref{tab:cons}, we can see that the proposed essential and disentangled constraints improve the quality of the memorized knowledge of the parametric knowledge base. Both the essential principle and the disentangled principle can solely contribute to the final performance, while the two complement each other and can collaborate for more qualified knowledge.

\subsubsection{Impact of the Knowledge Extractor}
\begin{figure}[h]
\centering  
\subfigure[AD.]{   
\begin{minipage}{4cm}
\centering    
\includegraphics[scale=0.5]{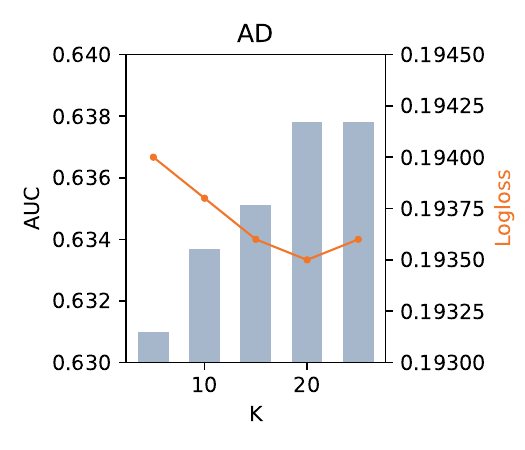}  
\end{minipage}
}
\subfigure[Eleme.]{ 
\begin{minipage}{4cm}
\centering    
\includegraphics[scale=0.5]{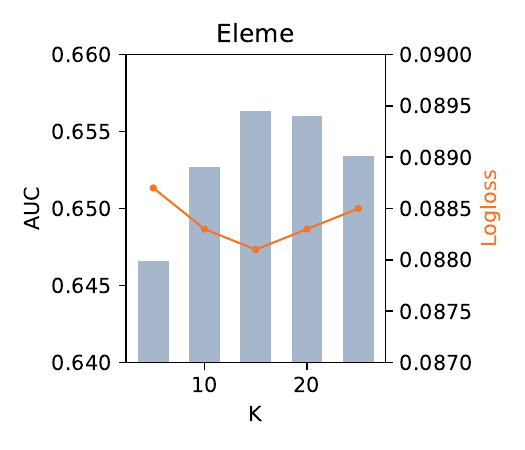}
\end{minipage}
}
\vspace{-10pt}
\caption{Performances of \modelname w.r.t different number of knowledge patterns per sample.}    
\label{fig:k}    
\end{figure}

The patterns of the old data space are diverse and complex, being hard to be fully memorized due to the existence of spurious patterns and entanglement.
To memorize the informative patterns within the data space, a knowledge extractor is designed to extract the essential and disentangled patterns within data points. 
In this section, we conduct experiments on the number of knowledge patterns per data point to study the impact of the knowledge extractor. Similarly, we select DeepFM as the backbone for AD dataset and DCN as the backbone for Eleme dataset. We test on the range of $\left[5,10,15,20,25\right]$, the results are listed in Figure~\ref{fig:k}. 
\vspace{-12pt}
\vspace{-10pt}

From the results, we can see that improving the number of knowledge vectors per data point generally gives rise to the performance, which validates the effectiveness of the knowledge extractor. In addition, further improving the number of knowledge patterns gives minor performance improvements or even impairs the final performance. This is because the increase of the knowledge patterns may result in redundancy and noise, which hurts the robustness of the model parameters.

\subsection{Case Study (RQ4)}

\subsubsection{Loss Curves of Different Objectives}
First of all, we study whether the losses corresponding to the two principles are learned well during the knowledge compression process. 
\vspace{-12pt}
\begin{figure}[h]
    \centering
    \includegraphics[width=0.46\textwidth]{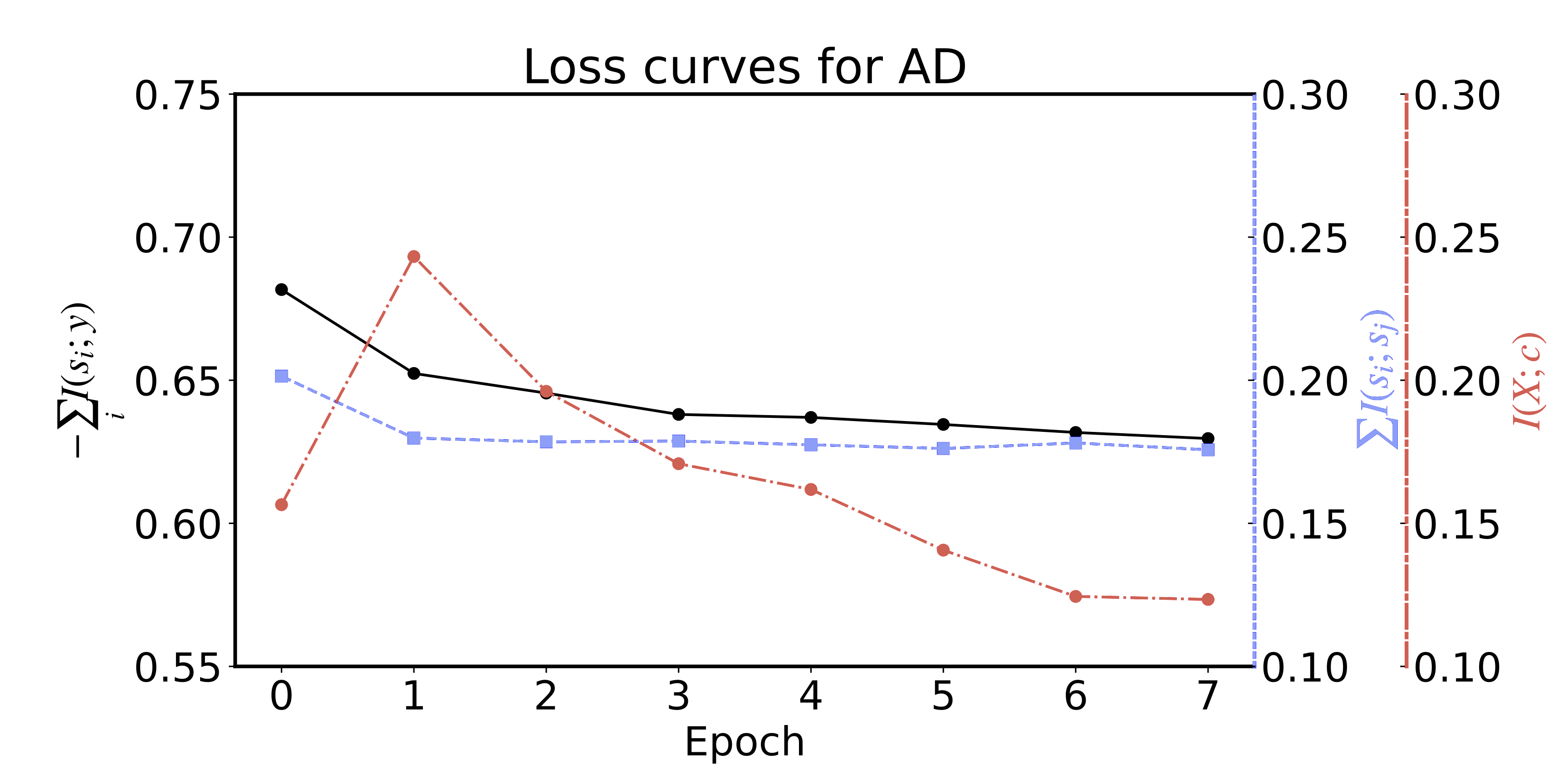}
    \vspace{-10pt}
    \caption{Learning curves of different losses on AD.}
    \label{fig:lossAD}
\end{figure}
\vspace{-10pt}

The learning curves of different losses on AD dataset is illustrated in Figure~\ref{fig:lossAD}. The blue line displays the loss curve for the disentangled constraint, which is $\sum_{i\neq j}I(s_i;s_j)$. The red line and black line together make up the loss for the essential constraint, which is $- \alpha\sum_j I(\mathbf{s_j};y)+ I(X;\mathbf{c})$. Concretely, the black line illustrates the curve for $-\sum_j I(\mathbf{s_j};y)$ which promotes the memorization for task relevant information, and the red line illustrates the curve for $I(X;\mathbf{c})$, which compresses the representation and helps to filter the noisy information. From the curves, we can see that all losses are learned well. While for the red one which corresponds to the vCLUB value, the MI upper bound estimation increases at the first epoch and decrease in the following training steps, since it takes some steps first to train the MI estimator well for correct estimation.

\subsubsection{Visualizations of the Extracted Patterns}
Then, we visualize the extracted patterns for more explicit analysis. 

\vspace{-10pt}
\begin{figure}[!h]
\centering  
\subfigure[AD]{
\label{Fig.sub.1}
\includegraphics[width=3.2cm,height = 4.5cm]{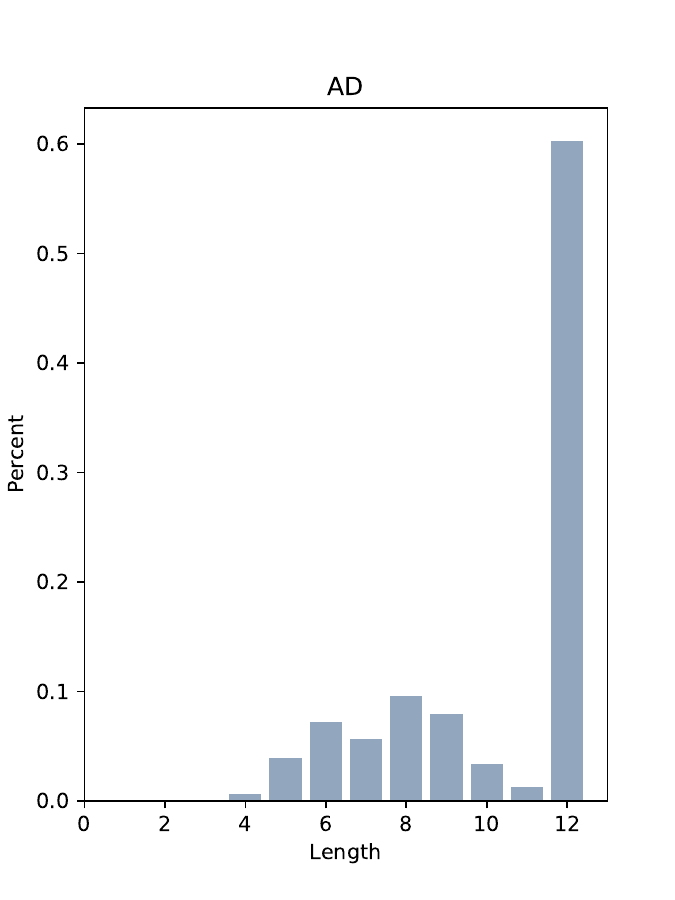}}
\subfigure[Eleme]{
\label{Fig.sub.2}
\includegraphics[width=4.8cm,height =4.5cm]{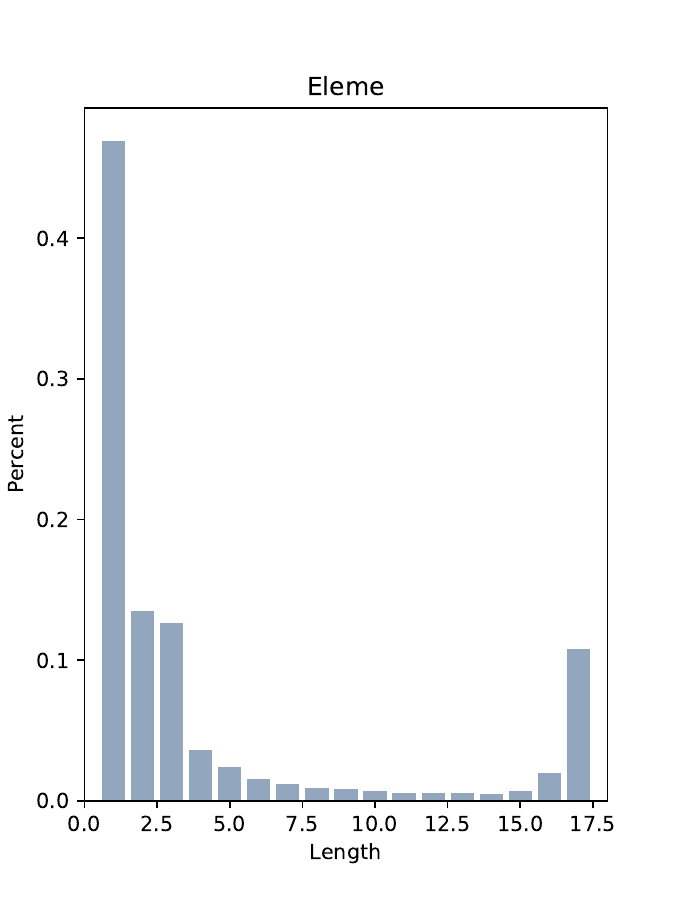}}
\vspace{-10pt}
\caption{Pattern scales. (Entry $>0.5$ is regarded as 1.)}
\label{illus}
\end{figure}
\vspace{-25pt}
\begin{figure}[h]
\centering  
\subfigure[AD]{
\label{Fig.sub.1}
\includegraphics[width=0.48\textwidth]{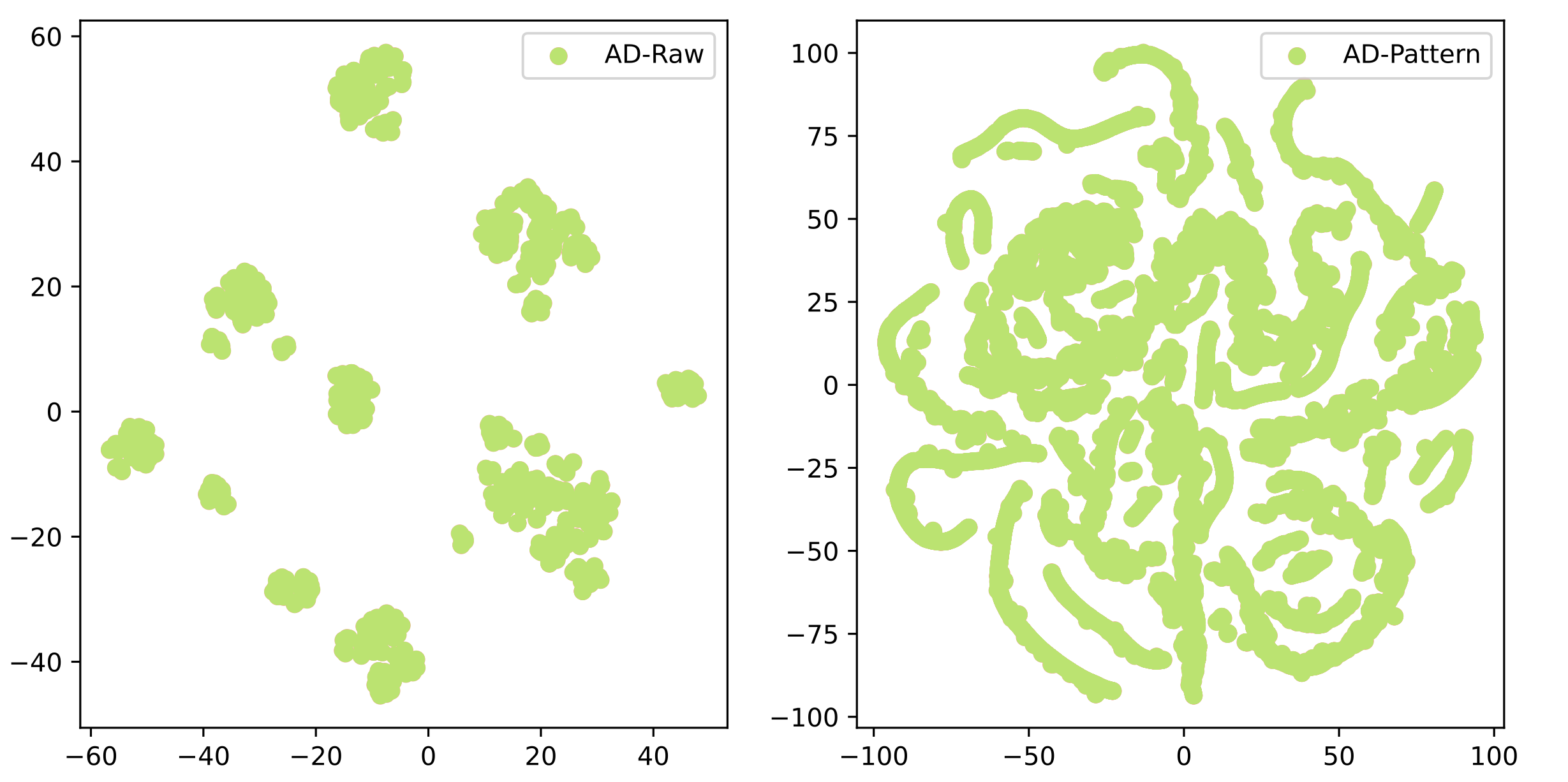}}
\subfigure[Eleme]{
\label{Fig.sub.2}
\includegraphics[width=0.48\textwidth]{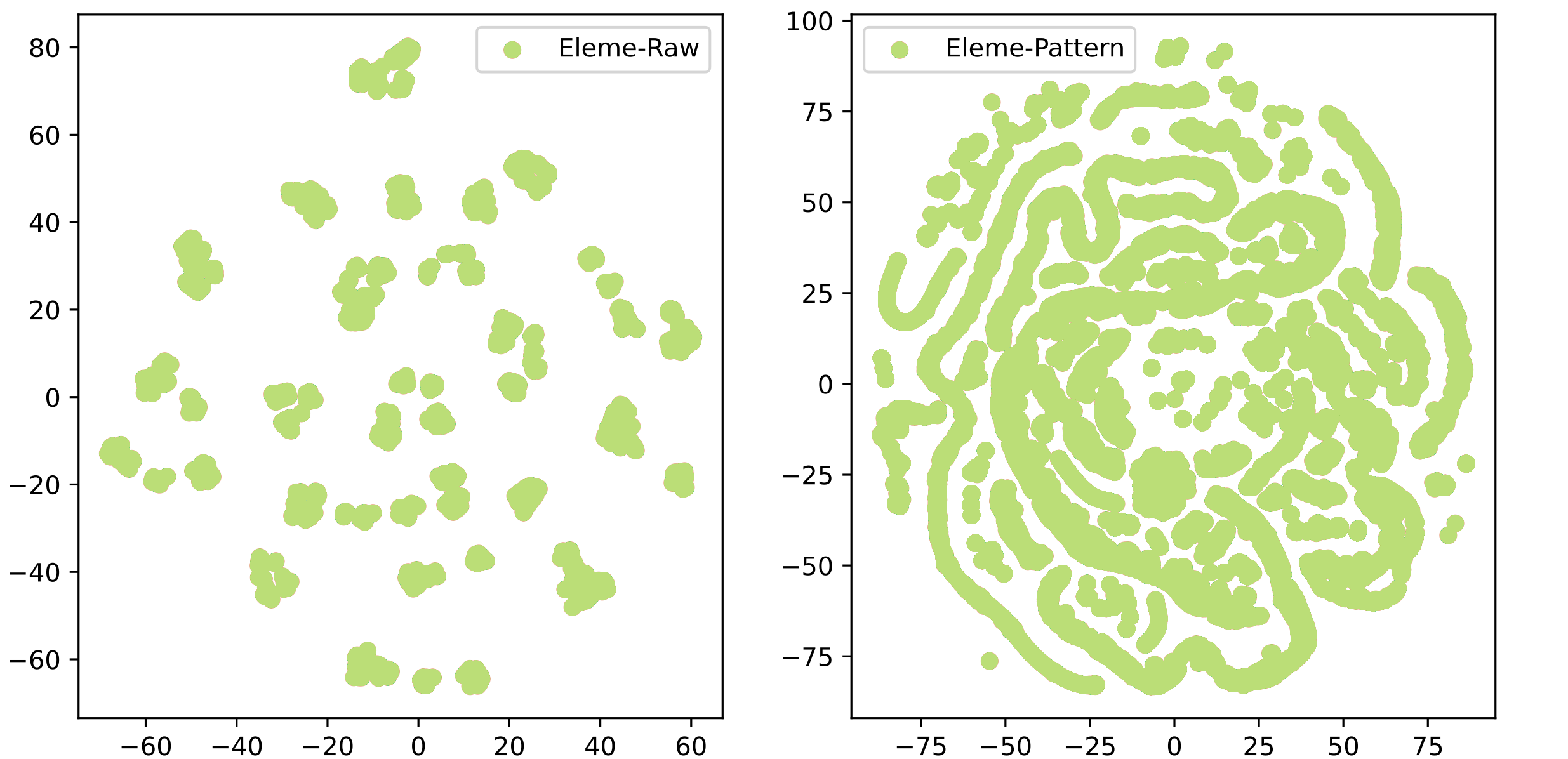}}
\vspace{-10pt}
\caption{T-SNE visualization for the representations of the raw data features and the extracted patterns. }
\label{tsne}
\end{figure}

Firstly, we study the masks generated by the knowledge extractor. Note that the entries of the masks are continuous. Here we regard an entry with value larger than $0.5$ as an existing feature and calculate the number of the non-zero entries of the generated masks. The extracted pattern scale distribution is illustrated in Figure~\ref{illus}. From the results, we could find that the scales of extracted patterns vary, which validates the effectiveness of the knowledge extractor for extracting patterns of arbitrary scales. 


In addition, we further study the representations of the raw data instance features and the extracted patterns. We use the T-SNE visualization to illustrate the distribution of the encodings of raw data instances and the extracted pattern vectors. The visualization results are displayed in Figure~\ref{tsne}.
Each point in the left denotes a representation of a raw data instance, and each point in the right represents an encoded pattern vector. 
From the results, we could see that there are apparently different manifolds existing in the extracted patterns representation space for both AD and Eleme datasets, while the raw data instances just appear in different clusterings. This validates the effectiveness of the proposed method on extracting the diverse and complex patterns from the old data.

\section{Conclusion}
In this paper, we design a parametric knowledge base \modelname that compresses the massive old data into compact knowledge stored in parameters.
Concretely, an instance-wise knowledge extractor is utilized to extract the patterns of arbitrary scales within data points and a knowledge encoder is utilized to memorize the patterns. 
The parametric knowledge base is regularized by the proposed essential and disentangled principles, which promote robust and generalized knowledge memorized by the parametric knowledge base without increasing the size of it.
\modelname is model-agnostic, flexible, and tailored for fine-grained instance-level knowledge augmentation. 

\begin{acks}
To Robert, for the bagels and explaining CMYK and color spaces.
\end{acks}

\bibliographystyle{ACM-Reference-Format}
\bibliography{sample-base}

\appendix
\section{Hyperparameter}
\label{sec:hyper}
We use consistent embedding size for all base models for fair comparison. For AD dataset, the embedding size for base models is set to 32. The size of the knowledge vectors for AD is set to 16. For Eleme dataset, the embedding size for base models is set to 16. The size of the knowledge vectors for Eleme is set to 8. The number of patterns to extract from each data instance is set to 20. The depth and number of attention heads for the knowledge encoder is 3 and 3, respectively. The learning rate is searched in the range of $[1e-4, 3e-4, 5e-4, 1e-3]$, while the weight decay is searched in the range of $[1e-4, 3e-4, 5e-4, 5e-5, 3e-5, 1e-5]$. Adam~\citep{kingma2014adam} is used for training.








\end{document}